\documentclass{article}

\usepackage{PRIMEarxiv}

\usepackage[utf8]{inputenc} 
\usepackage[T1]{fontenc}    
\usepackage{hyperref}       
\usepackage{url}            
\usepackage{booktabs}       
\usepackage{amsfonts}       
\usepackage{nicefrac}       
\usepackage{microtype}      
\usepackage{fancyhdr}       
\usepackage{graphicx}       
\usepackage{amsmath,amssymb,CJKutf8,enumitem,multirow,url}
\graphicspath{{media/}}     

\title{UtterTune: LoRA-Based Target-Language Pronunciation Edit and Control in Multilingual Text-to-Speech
}

\author{
  Shuhei KATO \\
  Independent Researcher \\
  Tokyo, Japan \\
  \texttt{shuhei@shuheikato.info} \\
}

\begin{document}
\maketitle

\begin{abstract}
We propose \textbf{UtterTune}, a lightweight method for adapting a multilingual text-to-speech (TTS) system built on a large language model (LLM). It improves control of pronunciation in the target language while preserving performance in the others. Although LLM architectures have enabled TTS models to achieve remarkable naturalness, accurately modeling grapheme-to-phoneme (G2P) mapping and prosody remains challenging, especially when the model omits an explicit G2P module and directly processes minimally encoded text (e.g., byte-pair encoding). UtterTune leverages low-rank adaptation to enable the control of segmental pronunciation and pitch accent at the phoneme level for Japanese speech, the target language in this paper, while maintaining naturalness and speaker similarity in a zero-shot setting. Objective and subjective evaluations confirm its effectiveness.
\end{abstract}


\section{Introduction}
\label{sec:intro}

Text-to-speech (TTS) models based on large language model (LLM) architecture (LLM-TTS in this paper) have demonstrated exceptional naturalness, especially in zero-shot multi-speaker and multilingual synthesis, leading the way in speech synthesis technology \cite{wang2023neuralcodeclanguagemodels,kharitonov-etal-2023-speak,chen2024valle2neuralcodec,du2024cosyvoice2scalablestreaming}; however, reproducing accurate pronunciation remains challenging. Some multilingual LLM-TTS, such as CosyVoice 2 \cite{du2024cosyvoice2scalablestreaming}, are designed to take raw text (characters) as input and tokenize it via byte-pair encoding (BPE) \cite{gage1994bpe}, without explicit phonemic or prosody markers. This design contrasts with conventional neural sequence-to-sequence TTS, which typically converts input text into phonemes (grapheme-to-phoneme; G2P) and prosody information, if needed, using a text frontend before feeding it into the model \cite{pyopenjtalk,jvsrecipe,vits,wang2023neuralcodeclanguagemodels}. Omitting the text frontend simplifies model architecture and facilitates multilingual training. On the other hand, because these models learn segmental pronunciation and prosody purely from data, they require a large number of speech-text pairs that cover the full diversity of the target language. This reliance on data fails for some languages---Japanese among them.

For example, the model occasionally cannot accurately predict Japanese segmental pronunciation and pitch accent, resulting in degraded speech quality and intelligibility. The reasons why data-driven G2P is challenging in Japanese include the fact that Japanese is written using a combination of \textit{kanji} (Chinese characters; ideograms), and \textit{kana} (phonograms), the use of more than 2,000 kanji in everyday language (joyo-kanji) \cite{joyokanjihyo}, and the inclusion of over 6,000 kanji in the Japanese Industrial Standards \cite{jisx0208}. Many kanji have two or more readings, and many specific combinations of kanji have their special readings. Since Japanese and Chinese share the same tokens for common kanji, Chinese pronunciations may appear in synthesized Japanese speech. From the perspective of pitch accent, accents are determined individually for each word, and the point of pitch falling (accent nuclei) may shift when words are linked \cite{nhkaccentdictionary,sagisaka1983accentuation}.

Users of conventional Japanese TTS can easily correct pronunciation errors by modifying the text frontend's output. But in models like CosyVoice 2, where G2P and accent estimation happen implicitly, users have no such handle. Segmental pronunciation can be corrected by using kana in the input text, but users cannot control the accent.

To solve this, we propose UtterTune. It provides intuitive pronunciation control for multilingual LLM-TTS without full fine-tuning or retraining from scratch. Our approach leverages low-rank adaptation (LoRA) \cite{hu2021loralowrankadaptationlarge}. LoRA is an efficient fine-tuning technique originally introduced for adapting LLMs by injecting small trainable weight matrices (low-rank updates) into the Transformer layers while keeping the original weights frozen. It drastically reduces the number of trainable parameters and avoids catastrophic forgetting of the base model’s capabilities. Several papers have applied LoRA to TTS for speaker adaptation and style control \cite{hsieh23_interspeech,louhaowei2024,kim24_interspeech,bondaruk2025lorpttslowrankpersonalizedtexttospeech}. UtterTune applies LoRA to the CosyVoice 2 model, with a specific focus on segmental pronunciation and pitch accent control in Japanese as the initial target. By fine-tuning only a small set of additional parameters on Japanese speech-transcription pair data, UtterTune enables editing and control to read input texts with accurate pronunciation, including native-like pitch accent. This approach does not affect the model’s performance on other languages because the LoRA module is not required.

Our contributions are summarized as follows: 1) We demonstrate the first application of LoRA to a multilingual LLM-TTS for language-specific segmental pronunciation and prosody edit and control, introducing a minimal intervention (only two special tokens, described later, and LoRA layers). 2) UtterTune preserves naturalness and speaker similarity in a zero-shot setting, as our objective and subjective evaluations show. 3) The adaptation is highly parameter-efficient: The LoRA updates form a small file, less than 0.5\% compared to the base model. The scripts and training/evaluation data are available at \url{https://github.com/shuheikatoinfo/UtterTune} for reproducibility; the pretrained LoRA weights are available at \url{https://huggingface.co/shuheikatoinfo/UtterTune-CosyVoice2-ja-JSUTJVS}; and audio samples are available on our demo page at \url{https://shuheikatoinfo.github.io/UtterTune}.

\section{Tokyo Japanese Pronunciation and Its Notation}

In this paper, we adopt Japanese, especially Tokyo Japanese, as the target language for the proposed method; therefore, we briefly describe its pronunciation and notation.

Segmental pronunciation of Tokyo Japanese speech can be written in kana, a set of phonograms. Kana is divided into \textit{hiragana} and \textit{katakana}. Both are completely interchangeable in terms of pronunciation. The number of hiragana or katakana characters is approximately 80 each \cite{jisx0208,utf8}. Most of the kana characters represent one mora. A single-mora sound can comprise multiple phonemes, e.g., a consonant and a subsequent vowel.

In terms of prosody, pitch accent is one of the most essential elements in Tokyo Japanese speech. In speech processing, pitch accent in Tokyo Japanese is often expressed as a binary value, indicating whether the mora is high (H) or low (L) \cite{sagisaka1983accentuation,openjtalk,pyopenjtalk}. A single accentual unit is called an accent phrase. An accent phrase contains at most one accent nucleus. Accent phrases generally begin with a low-high (LH) pattern, with the pitch falling from high to low, starting from the mora where the accent nucleus is located. As an exception, if the accent nucleus is at the first mora, accent phrases begin with a high-low (HL) pattern. Each word has its unique accent pattern, and there is a minimal tendency for which words have which accent patterns \cite{nhkaccentdictionary}. When words combine, the accent nucleus may shift, and there are shift patterns to some extent \cite{sagisaka1983accentuation}. Accents can help distinguish between words and can also indicate semantic cohesion \cite{nhkaccentdictionary,sagisaka1983accentuation}.

A standard notation for Tokyo Japanese pronunciation exists for speech synthesis \cite{jeitait4006}. In this standard, segmental pronunciation can be represented using katakana, with accent nuclei indicated by apostrophes (’) and accent phrase boundaries indicated by slashes (/).

\section{Proposed Method}
\label{sec:proposed_method}

\subsection{Base Model}
\label{sec:pagestyle}

We build UtterTune on CosyVoice 2, a state-of-the-art multilingual LLM-TTS system. CosyVoice 2 treats speech as discrete tokens encoded by a semantic codec (a vector quantizer). During synthesis, a unified text-speech language model (LM; Qwen2.5-0.5B \cite{hui2024qwen25codertechnicalreport}) generates these tokens from input text. The generated tokens are then passed to a vocoder to produce a waveform. Text input is processed by a BPE tokenizer (shared across languages) instead of a conventional phoneme-based text front-end. This design enables the model to operate end-to-end on raw text. On the other hand, it also forces the model to learn pronunciation from data. The CosyVoice 2 LM is trained in an autoregressive next-token prediction manner. The LM treats the text and subsequent speech tokens in a single sequence. While effective in many cases, this approach has limitations for Japanese, as described in the introduction. Indeed, the CosyVoice 2 paper reports that the model’s performance on Japanese lags behind that of other languages (e.g., English, Chinese) \cite{du2024cosyvoice2scalablestreaming}.

\subsection{LoRA Integration in Transformer Layers}
\label{sec:typestyle}

To add pronunciation controllability of Japanese to CosyVoice 2 without full fine-tuning, we utilize LoRA \cite{hu2021loralowrankadaptationlarge}. LoRA inserts trainable low-rank matrices into the Transformer’s weight structure, typically by decomposing the large weight matrices of each layer’s query, key, value, and output (Q, K, V, and O) projections in the multi-head self-attention mechanism. We freeze all the original weights of CosyVoice 2 and add LoRA layers to every self-attention block in the text-speech LM. Concretely, for each dense weight matrix $W \in \mathbb{R}^{d \times k}$ in the Q/K/V/O linear transformations, we introduce a pair of smaller matrices $B$ and $C$ of rank $r$ (with $r \ll \min(d,k)$). During training, the effective weight becomes $W + \alpha \cdot B C$, where $\alpha$ is a scaling factor and $B C$ has rank $r$. Only the $B$ and $C$ parameters and embeddings for new tokens, as described in the following subsection, are updated during training. LoRA dramatically reduces the number of trainable parameters, less than 0.5\% of the base model. This parameter efficiency not only saves GPU memory and training time, but also tends to preserve the base model’s knowledge, since the pre-trained weights remain untouched.

\subsection{Phoneme Inputs with Special Tokens}
\label{subsec:phoneme_inputs_with_special_tokens}

Another key component of UtterTune is intuitive input that guides the model’s segmental pronunciation and pitch accent. CosyVoice 2 takes raw Japanese text (kanji, kana, etc.) as input. Such raw text does not always make segmental pronunciation and pitch accent explicit.

We extend the model’s input representation by adding two special tag tokens, \texttt{<PHON\_START>} and \texttt{<PHON\_END>}, to the vocabulary. When the model encounters a \texttt{<PHON\_START>} token, it indicates that the subsequent sequence represents phonemes (segmental pronunciation; regarding kanas as phonemes here for simplicity) and accent rather than ordinary text, until the \texttt{<PHON\_END>} token is reached. We fine-tune the model such that it can accept inputs of the form: e.g., \texttt{<PHON\_START>}\begin{CJK}{UTF8}{ipxm}チ'ミ/モーリョー\end{CJK}\texttt{<PHON\_END>} to synthesize speech of ``\begin{CJK}{UTF8}{ipxm}魑魅魍魎\end{CJK}'' (``evil spirits and monsters''). For the model training, following the standard \cite{jeitait4006}, we convert a portion of the input sentences to katakana, with accent nuclei indicated by apostrophes (’) and accent phrase boundaries indicated by slashes (/), and wrap them in the new tag tokens. This teaches the model to map phonemic sequences with accent information to the correct speech tokens. We keep the original embedding matrix frozen. Only the embeddings of the new tag tokens, which are randomly initialized, are updated during fine-tuning. This approach is analogous to adding a ``mode switch'' for the model: In normal mode (no tags), it is expected to read text as before; in phoneme mode, it follows a supplied spelling.

\textbf{Note on precedence}. The mechanism proposed here---delimiting a target span of the input text with two dedicated special tokens and replacing it with a kana-plus-pitch-accent notation, then fine-tuning an LLM-TTS backbone to honor it at inference time---was first released in August 2025 (arXiv v1 of this paper \cite{kato2025uttertune}). A construction essentially identical to this was subsequently presented as the PronSteering mechanism in the Sarashina2.2-TTS technical report \cite{liu2026sarashina2.2tts}, published in June 2026, which likewise introduces two special tokens (\texttt{<|pron\_start|>}, \texttt{<|pron\_end|>}) and replaces the target text with a kana reading and pitch-accent tags following another prosodic notation \cite{kurihara2021prosodicfeaturescontrol}. We note this correspondence to record the precedence of the formulation introduced in this work.

\section{Experiments}
\label{sec:experiments}

\subsection{Training and Validation Data}
\label{ssec:training_validation_data}

For adaptation, we used two public speech corpora: JSUT \cite{sonobe2017jsutcorpusfreelargescale}, a female single-speaker corpus, and JVS \cite{takamichi2019jvscorpusfreejapanese}, a multi-speaker corpus comprising 100 speakers. We utilized the ``basic5000'' and ``voiceactress100'' sets from JSUT, and the ``parallel100'' set from JVS. One randomly selected noun from each transcript was converted into a phonemic sequence for the \texttt{<PHON\_START>} \dots \texttt{<PHON\_END>} format described in \ref{subsec:phoneme_inputs_with_special_tokens}, using pyopenjtalk \cite{pyopenjtalk}, a publicly available Japanese G2P tool. The author, a native speaker, then modified estimation errors.

The training process thus pairs these transcripts with the original speech recordings. As a result, the dataset comprised 15,097 Japanese speech-transcription pairs. We held out 754 utterances from it for validation.

\subsection{Evaluation Data}
\label{ssec:evaluation_data}

For evaluation, we prepared two test sets: 1) a general test set and 2) a targeted segmental pronunciation and pitch accent controllability stress-test set.

Test Set 1 consisted of 50 Japanese sentences, each 20 to 40 characters in length, including punctuation. The sentences were obtained by instructing ChatGPT o3 \cite{chatgpto3} to generate easy-to-read Japanese sentences because CosyVoice 2 sometimes mispronounces sentences in public speech corpora. After the generation, the author reviewed all the generated sentences and made necessary grammatical or semantic modifications. Using these sentences as input, we performed zero-shot synthesis using prompt audio and transcription derived from ReazonSpeech v2.0 \cite{reazonspeech}, a 35,000-hour Japanese speech corpus collected from television broadcasts. We chose 48 audio clips, each representing one speaker, with a signal-to-noise ratio (SNR) estimated by WADA-SNR \cite{kim2008robust} of 25\,dB or higher and lasting 4 seconds or less. We then performed voice activity detection (VAD) using the \texttt{torchaudio.functional.vad} module \cite{torchaudio.functional.vad} on these audio clips and used 42 audio clips and their transcription as the prompts for which speech segments were detected. We also applied the same VAD to the synthesized speech because we observed that CosyVoice 2 occasionally generates non-speech sounds before or after speech segments. As a result, we obtained 2,100 stimuli per system.

Test Set 2 consisted of 50 sentences designed to assess the controllability of segmental pronunciation and pitch accent. The sentences were obtained by instructing ChatGPT o3 to generate easy-to-read Japanese sentences that include one difficult-to-read word each because no appropriate public benchmarks existed. After the generation, the author also reviewed all the generated sentences and made necessary modifications. The difficult words lasted from 2 to 6 morae, and the accent nuclei were located from the first to the last morae. Accent annotations followed the NHK Japanese Accent Dictionary \cite{nhkaccentdictionary} for the words where the entry existed. For the proposed method, the pronunciation of the target difficult-to-read word was spelled out in the \texttt{<PHON\_START>} \dots \texttt{<PHON\_END>} format. We performed zero-shot synthesis on the same speakers used in Test Set 1. As a result, we obtained 2,100 stimuli per system.

\subsection{Fine-Tuning Configuration}
\label{sec:print}

We adopted the pre-trained CosyVoice 2-0.5B checkpoint \cite{cosyvoice2checkpoint} and applied our LoRA modifications. The LoRA rank, scaling alpha, and dropout rate were set to 16, 64, and 0.05, respectively, for all injected layers. We trained using the AdamW \cite{loshchilov2019adamw} optimizer with a learning rate of $1 \times 10^{-4}$. We used a warm-up schedule for the first 10\% of training steps and cosine decay. The mini-batch size was set to 8, and we fine-tuned for 20,000 steps on an NVIDIA RTX 4090 GPU in under 1 hour.

The only parameters updated were the LoRA layers and the embeddings for \texttt{<PHON\_START>} and \texttt{<PHON\_END>}. We did not apply any auxiliary losses; the training objective remained cross-entropy for next-token prediction of speech tokens given the text prompt.

\subsection{Baseline Systems}
\label{subsec:baseline_systems}

We employed the official CosyVoice 2-0.5B checkpoint \cite{cosyvoice2checkpoint} as the baseline. For Test Set 2, we added a kana‑input baseline, replacing the transcription of the target difficult-to-read word with katakana in the input texts.

\subsection{Evaluation Metrics}
\label{subsec:evaluation_metrics}

\begin{description}[leftmargin=0pt]
    \item[{\parbox[b]{\linewidth}{Automatic naturalness mean opinion score (MOS) prediction}}]
    We utilized UTMOSv2, a non-intrusive model that predicts the MOS of a given speech sample \cite{baba2024utmosv2}. Samples with UTMOSv2 less than 2.0 in any system were excluded from the tabulation to exclude the effect of extremely inferior quality samples due to synthesis errors. The ratio of dropped samples was less than 2\% for each test set and system.
    \item[{\parbox[b]{\linewidth}{Naturalness MOS}}]
    We conducted a subjective listening test in which participants listened to synthesized speech and rated its naturalness on a 5-point MOS scale. 16 crowdworkers who are native Japanese speakers participated in this test. We gathered MOS scores for 126 stimuli, paired for each model and stratified by UTMOSv2 scores from Test Set 1, with a minimum of 4 ratings per sample. The stimuli were selected to contain all 42 speakers.
    \item[{\parbox[b]{\linewidth}{Objective speaker similarity (SS)}}]
    Although our focus is on pronunciation controllability, we want to ensure the adapted model does not degrade the ability to maintain the target speaker’s voice characteristics. We measure speaker similarity using ERes2Net \cite{chen2023eres2net,eres2netcheckpoint}. Speaker embeddings were extracted from synthesized speech and the corresponding prompt speech of the same speaker. We then compute the cosine similarity between these embeddings.
    \item[{\parbox[b]{\linewidth}{Character error rate (CER)}}]
    To quantify intelligibility, we used an automatic speech recognition (ASR) system to transcribe the synthesized speech and calculated CER against the input text. We employed OpenAI’s Whisper large-v3 \cite{whisperlargev3} to transcribe the outputs. Since Test Set 2 contains difficult-to-read words, Whisper was forced to output only kana and a limited set of symbols (U+3000--U+30FF in Unicode \cite{unicode}). We then removed all spaces and symbols from the transcription and converted it into katakana for CER calculation\footnote{Evaluating reading correctness by transcribing synthesized speech into kana and computing a CER was adopted in the initial release of this work \cite{kato2025uttertune}. We note that the Sarashina2.2-TTS \cite{liu2026sarashina2.2tts} later developed this evaluation into a dedicated kana-output ASR model and released it, together with a joyo-kanji reading benchmark and evaluation scripts, as open-source resources \cite{kana-whisper,joyo-kanji-yomi-benchmark-dataset,joyo-kanji-yomi-benchmark-scripts}; such reproducible tooling is a valuable contribution to the community, and future work on Japanese pronunciation evaluation would benefit from building on it. Motivated by them, we also release the training data (kana-plus-accent-tagged pairs derived from JSUT and JVS) and the three evaluation sets (difficult / katakana / easy) used in this work at \url{https://github.com/shuheikatoinfo/UtterTune}.}. The same procedure was used for Test Set 1. Some prompt speech has been observed to leak to the beginning of the synthesized speech, and the leakage portion was excluded from the CER calculation. Samples with CER $>0.5$ were omitted from aggregation to avoid serious ASR errors.
    \item[{\parbox[b]{\linewidth}{Accent correctness}}]
    We conducted another subjective listening test in which participants listened to synthesized speech and identified whether the target word's pitch accent was spoken as instructed. 15 crowdworkers who are native Japanese speakers and passed a check test\footnote{It is known that even native Japanese speakers find it challenging to accurately identify the Tokyo Japanese accent when listening.} participated in this test. We gathered ratings for 126 stimuli paired for each model, stratified by sentence number from Test Set 2, with a minimum of two ratings per sample. The stimuli were selected to contain all 42 speakers.
\end{description}

\section{Results and Discussion}
\label{sec:results_and_discussion}

\subsection{Naturalness and Speaker Similarity}

\begin{table*}[t]
    \centering
    \begin{tabular}{clrrrrr}
        \toprule
        \textbf{Test Set} & \textbf{System} & \textbf{UTMOSv2}$\uparrow$ & \textbf{MOS}$\uparrow$ & \textbf{SS}$\uparrow$ & \textbf{CER}$\downarrow$ & \textbf{Accent correctness}$\uparrow$\\
        \midrule
        \multirow{2}{*}{1} & CosyVoice 2-0.5B (baseline) & 3.25 & 3.44 & 0.693 & \textbf{0.0557} & -\\
        & Proposed & \textbf{3.28} & \textbf{3.88} & 0.696 & 0.0639 & -\\
        \midrule
        \multirow{3}{*}{2} & CosyVoice 2-0.5B (baseline) & 3.27 & - & 0.695 & 0.1005 & -\\
        & CosyVoice 2-0.5B (kana input) & 3.28 & - & 0.692 & \textbf{0.0595} & 0.498\\
        & Proposed & 3.30 & - & 0.697 & 0.0626 & \textbf{0.899}\\
        \bottomrule
    \end{tabular}
    \caption{Experimental results. Bold text indicates values that were significantly ($p<0.05$) better for each test set based on the one-sided paired Wilcoxon test (and Bonferroni correction if needed).}
    \label{tab:experimental_results}
\end{table*}

Table \ref{tab:experimental_results} summarizes the objective and subjective scores. On Test Set 1, UtterTune increased the crowd-sourced MOS from 3.44 to 3.88. A one‑sided paired Wilcoxon test confirmed the gain was statistically significant ($p<0.0001$). Taken together, the UTMOSv2 results show that UtterTune does not degrade naturalness.

For Test Set 2, UTMOSv2 scores showed no significant differences among systems, and the 99\% confidence interval (CI) for the difference scores was less than 0.06 for all system pairs. The three systems can be considered equivalent.

Across both test sets, the ERes2Net cosine similarity hovered around 0.69 for all systems and showed no significant differences. This means that UtterTune kept the ability to maintain the voice characteristics in the prompt speech.

\subsection{Effect of Phoneme Tags and Kana Transcription}

Comparing the results of the baseline and kana input systems, feeding kana transcription alone reduced CER from 0.1005 to 0.0595. CosyVoice 2 could pronounce words segmentally correctly once kanji ambiguity was resolved. Yet its accent correctness was low (0.498): the model often read the proper phonemes with an incorrect pitch pattern.

UtterTune, by contrast, attained 0.899 accent correctness while keeping a competitive CER (0.0626). The difference of CER to the kana-baseline was significant ($p<0.005$), but the effect size was negligible (Cliff's delta$= 0.115$ \cite{cliff1993}). The LoRA adaptation learned not only to obey the given phoneme string but also to place the pitch fall at the right mora.

To confirm that the effect was not extended to the parts not enclosed by \texttt{<PHON\_START>} and \texttt{<PHON\_END>}, an additional 50 sentences containing only one katakana word with an accented nucleus were prepared based on the ChatGPT o3 generation. Accent annotations followed the NHK Japanese Accent Dictionary \cite{nhkaccentdictionary}. If affected, these words should be synthesized with a flat pitch pattern. The sentences were synthesized by all 42 speakers using the baseline and the proposed method. The percentage of correct accents, measured using the same procedure as the accent correctness, was 0.796 (baseline) and 0.829 (proposed method), with no significant difference. This means that UtterTune did not leak pronunciation controllability outside the tag tokens.

\section{Conclusion}
\label{sec:conclusion}

We proposed UtterTune, a lightweight LoRA-based adaptation that endows a multilingual LLM-TTS model with phoneme-level pronunciation control. By freezing the 0.5-B-parameter CosyVoice 2 backbone and fine‑tuning only rank‑16 LoRA matrices plus two special input tokens, UtterTune requires less than 0.5\% additional parameters and a few GPU‑hours of training. It raised naturalness from 3.44 to 3.88 MOS, kept CER below 7\% on everyday sentences, and boosted accent correctness from 0.498 to 0.899 on stress‑test material without degrading speaker similarity. In principle, UtterTune never degrades in performance across other languages because LoRA does not need to be applied during synthesis. These results demonstrate that minor, targeted updates can rectify language-specific deficiencies—here, Japanese G2P and pitch accent—while preserving the strong performance of a multilingual model.

Future research includes applying UtterTune to other Japanese dialects and other languages, as well as verifying its performance during code switching.

\bibliographystyle{unsrturl}
\bibliography{refs}

@inproceedings{baba2024utmosv2,
    title={The {T05} System for The {VoiceMOS Challenge} 2024: Transfer Learning from Deep Image Classifier to Naturalness {MOS} Prediction of High-Quality Synthetic Speech},
    author={Baba, Kaito and Nakata, Wataru and Saito, Yuki and Saruwatari, Hiroshi},
    booktitle={Proc. IEEE Spoken Lang. Technol. Workshop},
    year={2024},
}

@misc{bondaruk2025lorpttslowrankpersonalizedtexttospeech,
    title={{LoRP-TTS}: Low-Rank Personalized Text-To-Speech}, 
    author={{\L}ukasz Bondaruk and Jakub Kubiak},
    year={2025},
    howpublished={arXiv:2502.07562 [cs.SD]},
}

@inproceedings{chen2023eres2net,
    author={Yafeng Chen and Siqi Zheng and Hui Wang and Luyao Cheng and Qian Chen and Jiajun Qi},
    title={An Enhanced {Res2Net} with Local and Global Feature Fusion for Speaker Verification},
    booktitle={Proc. INTERSPEECH},
    year={2023}, 
}

@misc{chen2024valle2neuralcodec,
    title={{VALL-E 2}: Neural Codec Language Models are Human Parity Zero-Shot Text to Speech Synthesizers}, 
    author={Sanyuan Chen and Shujie Liu and Long Zhou and Yanqing Liu and Xu Tan and Jinyu Li and Sheng Zhao and Yao Qian and Furu Wei},
    year={2024},
    howpublished={arXiv:2406.05370 [cs.CL]},
}

@misc{chatgpto3,
    author={{OpenAI}},
    title={{OpenAI} o3 and o4-mini},
    url={https://help.openai.com/en/articles/9624314-model-release-notes#h_cfb262abff},
}

@article{cliff1993,
    author={Norman Cliff},
    title={Dominance statistics: Ordinal analyses to answer ordinal questions},
    journal={Psychological Bull.},
    volume={114},
    number={3},
    pages={494--509},
    year={1993},
}

@misc{cosyvoice2checkpoint,
    author={{Inst. Intell. Comput.}},
    title={{CosyVoice2-0.5B}},
    url={https://www.modelscope.cn/models/iic/CosyVoice2-0.5B},
}

@misc{du2024cosyvoice2scalablestreaming,
    title={{CosyVoice} 2: Scalable Streaming Speech Synthesis with Large Language Models}, 
    author={Zhihao Du and Yuxuan Wang and Qian Chen and Xian Shi and Xiang Lv and Tianyu Zhao and Zhifu Gao and Yexin Yang and Changfeng Gao and Hui Wang and Fan Yu and Huadai Liu and Zhengyan Sheng and Yue Gu and Chong Deng and Wen Wang and Shiliang Zhang and Zhijie Yan and Jingren Zhou},
    year={2024},
    howpublished={arXiv:2412.10117 [cs.SD]},
}

@misc{eres2netcheckpoint,
    title={iic/speech\_eres2net\_sv\_zh-cn\_16k-common},
    url={https://www.modelscope.cn/models/iic/speech_eres2net_sv_zh-cn_16k-common}
}

@article{gage1994bpe,
    author={Gage, Philip},
    title={A new algorithm for data compression},
    year={1994},
    volume={12},
    number={2},
    issn={0898-9788},
    journal={C Users J.},
    pages={23--38},
    numpages={16}
}

@inproceedings{hsieh23_interspeech,
    title={Adapter-Based Extension of Multi-Speaker Text-To-Speech Model for New Speakers},
    author={Cheng-Ping Hsieh and Subhankar Ghosh and Boris Ginsburg},
    year={2023},
    booktitle={Proc. INTERSPEECH},
    pages={3028--3032},
}

@misc{hui2024qwen25codertechnicalreport,
    title={Qwen2.5-Coder Technical Report}, 
    author={Binyuan Hui and Jian Yang and Zeyu Cui and Jiaxi Yang and Dayiheng Liu and Lei Zhang and Tianyu Liu and Jiajun Zhang and Bowen Yu and Keming Lu and Kai Dang and Yang Fan and Yichang Zhang and An Yang and Rui Men and Fei Huang and Bo Zheng and Yibo Miao and Shanghaoran Quan and Yunlong Feng and Xingzhang Ren and Xuancheng Ren and Jingren Zhou and Junyang Lin},
    year={2024},
    howpublished={arXiv:2409.12186 [cs.CL]},
}

@misc{hu2021loralowrankadaptationlarge,
    title={{LoRA}: Low-Rank Adaptation of Large Language Models}, 
    author={Edward J. Hu and Yelong Shen and Phillip Wallis and Zeyuan Allen-Zhu and Yuanzhi Li and Shean Wang and Lu Wang and Weizhu Chen},
    year={2021},
    howpublished={arXiv:2106.09685 [cs.CL]}
}

@article{kharitonov-etal-2023-speak,
    title = "Speak, Read and Prompt: High-Fidelity Text-to-Speech with Minimal Supervision",
    author = {Kharitonov, Eugene  and
      Vincent, Damien  and
      Borsos, Zal{\'a}n  and
      Marinier, Rapha{\"e}l  and
      Girgin, Sertan  and
      Pietquin, Olivier  and
      Sharifi, Matt  and
      Tagliasacchi, Marco  and
      Zeghidour, Neil},
    journal = "Trans. Assoc. Comput. Linguistics",
    volume = "11",
    year = "2023",
    pages = "1703--1718",
}

@inproceedings{kim24_interspeech,
  title={{VoiceTailor}: Lightweight Plug-In Adapter for Diffusion-Based Personalized Text-to-Speech},
  author={Heeseung Kim and Sang-gil Lee and Jiheum Yeom and Che Hyun Lee and Sungwon Kim and Sungroh Yoon},
  year={2024},
  booktitle={Proc. INTERSPEECH},
  pages={4413--4417},
}

@misc{jeitait4006,
    title={{JEITA IT-4006}},
    author={{Jpn. Electronics and Inf. Technol. Industries Assoc.}},
    year={2010},
}

@misc{jisx0208,
    title={{JIS X0208}},
    author={{Minister of Int. Trade and Industry}},
    year={1978},
}

@misc{joyokanjihyo,
    title={Joyo Kanji-hyo},
    author={Prime Minister of Jpn.},
    year={2010},
    url={https://www.bunka.go.jp/kokugo_nihongo/sisaku/joho/joho/kijun/naikaku/pdf/joyokanjihyo_20101130.pdf},
}

@misc{jvsrecipe,
    title={{JVS RECIPE}},
    author={Tomoki Hayashi and others},
    year={2020},
    url={https://github.com/espnet/espnet/tree/master/egs2/jvs/tts1},
}

@inproceedings{kim2008robust,
    title={Robust signal-to-noise ratio estimation based on waveform amplitude distribution analysis},
    author={Kim, Chanwoo and Stern, Richard M},
    booktitle={Proc. INTERSPEECH},
    pages={2598--2601},
    year={2008}
}

@inproceedings{loshchilov2019adamw,
    author={Ilya Loshchilov and Frank Hutter},
    title={Decoupled Weight Decay Regularization},
    booktitle={Proc. Int. Conf. Learn. Representations},
    year={2019},
}

@inproceedings{louhaowei2024,
    author={Lou, Haowei and Paik, Hye-Young and Hu, Wen and Yao, Lina},
    title={{StyleSpeech}: Parameter-efficient Fine Tuning for Pre-trained Controllable Text-to-Speech},
    booktitle={Proc. ACM Int. Conf. Multimedia in Asia},
    year={2024},
}

@book{nhkaccentdictionary,
    editor={{NHK Broadcast. Culture Res. Inst.}},
    title={NHK's New Dictionary of Japanese Pronunciation and Accentuation},
    publisher={NHK Publishing},
    year={2016},
}

@misc{openjtalk,
    title={{Open JTalk version 1.11}},
    author={Keiichi Tokuda and Keiichio Oura and Shinji Sako},
    year={2018},
    url={https://open-jtalk.sourceforge.net/},
}

@misc{pyopenjtalk,
    title={{pyopenjtalk}},
    author={Ryuichi Yamamoto and others},
    year={2018},
    url={https://github.com/r9y9/pyopenjtalk},
}

@misc{reazonspeech,
    title={reazonspeech},
    author={{Reazon Human Interaction Lab}},
    year={2024},
    url={https://huggingface.co/datasets/reazon-research/reazonspeech},
}

@article{sagisaka1983accentuation,
    author={Yoshinori Sagisaka and Hirokazu Sato},
    title={Accentuation Rules for {Japanese} Word Concatenation},
    volume={J66-D},
    number={7},
    journal={The IEICE Trans. (Japanese Ed.)},
    year={1983},
    pages={849--856},
}

@misc{kato2025uttertune,
    title={{UtterTune}: {LoRA}-Based Target-Language Pronunciation Edit and Control in Multilingual Text-to-Speech},
    author={Shuhei Kato},
    year={2025},
    howpublished={arXiv:2508.09767v1 [cs.SD]},
}

@misc{liu2026sarashina2.2tts,
    title={{Sarashina2.2-TTS}: Tackling Kanji Polyphony in {Japanese} Speech Generation via Data Scaling and Targeted Data Synthesis},
    author={Lianbo Liu and Shiao Zhu and Kai Washizaki and Reo Yoneyama and Haesung Jeon and Mengjie Zhao and Yusuke Fujita and Hao Shi and Nao Yoshida and Yuan Gao and Roman Koshkin and Yukiya Hono and Yui Sudo},
    year={2026},
    howpublished={arXiv:2606.25369 [cs.SD]},
}

@article{kurihara2021prosodicfeaturescontrol,
    author={Kiyoshi Kurihara and Nobumasa Seiyama and Tadashi Kumano},
    title={Prosodic Features Control by Symbols as Input of
Sequence-to-Sequence Acoustic Modeling for Neural {TTS}},
    journal={IEICE Trans. Inf. \& Syst.},
    volume={E104-D},
    number={2},
    year={2021},
}

@misc{sonobe2017jsutcorpusfreelargescale,
    title={{JSUT} corpus: Free large-scale {Japanese} speech corpus for end-to-end speech synthesis}, 
    author={Ryosuke Sonobe and Shinnosuke Takamichi and Hiroshi Saruwatari},
    year={2017},
    howpublished={arXiv:1711.00354 [cs.CL]}, 
}

@misc{takamichi2019jvscorpusfreejapanese,
    title={{JVS} corpus: Free {Japanese} multi-speaker voice corpus}, 
    author={Shinnosuke Takamichi and Kentaro Mitsui and Yuki Saito and Tomoki Koriyama and Naoko Tanji and Hiroshi Saruwatari},
    year={2019},
    howpublished={arXiv:1908.06248 [cs.SD]}, 
}

@misc{torchaudio.functional.vad,
    key={torchaudio.functional.vad},
    title={torchaudio.functional.vad},
    year={2024},
    url={https://docs.pytorch.org/audio/2.7.0/generated/torchaudio.functional.vad.html},
}

@misc{unicode,
    author={{Unicode, Inc.}},
    title={Unicode® 16.0.0},
    url={https://www.unicode.org/versions/Unicode16.0.0/}, 
}

@misc{utf8,
    title={{ISO/IEC 10646:2020/Amd 1:2023}},
    author={{Int. Org. Standardization}},
    year={2023},
}

@misc{kana-whisper,
    author={{SB Intuitions}},
    title={{kana-whisper}},
    year={2026},
    url={https://huggingface.co/sbintuitions/kana-whisper},
}

@misc{joyo-kanji-yomi-benchmark-dataset,
    author={{SB Intuitions}},
    title={{joyo-kanji-yomi-benchmark}},
    year={2026},
    url={https://huggingface.co/datasets/sbintuitions/joyo-kanji-yomi-benchmark},
}

@misc{joyo-kanji-yomi-benchmark-scripts,
    author={{SB Intuitions}},
    title={{Joyo-Kanji-Yomi-Benchmark}},
    year={2026},
    url={https://github.com/sbintuitions/Joyo-Kanji-Yomi-Benchmark},
}

@misc{wang2023neuralcodeclanguagemodels,
    title={Neural Codec Language Models are Zero-Shot Text to Speech Synthesizers}, 
    author={Chengyi Wang and Sanyuan Chen and Yu Wu and Ziqiang Zhang and Long Zhou and Shujie Liu and Zhuo Chen and Yanqing Liu and Huaming Wang and Jinyu Li and Lei He and Sheng Zhao and Furu Wei},
    year={2023},
    howpublished={arXiv:2301.02111 [cs.CL]},
}

@misc{whisperlargev3,
    author={{OpenAI}},
    title={whisper-large-v3},
    year={2023},
    url={https://huggingface.co/openai/whisper-large-v3},
}

@inproceedings{vits,
    title={Conditional Variational Autoencoder with Adversarial Learning for End-to-End Text-to-Speech},
    author={Jaehyeon Kim and Jungil Kong and Juhee Son},
    booktitle={Proc. Int. Conf. Mach. Learn.},
    year={2021},
}

\end{document}